# Control of Regional and Global Weather

## Alexander Bolonkin

C&R, 1310 Avenue R, #F-6, Brooklyn, NY 11229, USA
T/F 718-339-4563, aBolonkin@juno.com, http://Bolonkin.narod.ru

## Abstract

Author suggests and researches a new revolutionary idea for regional and global weather control. He offers to cover cities, bad regions of country, full country or a continent by a thin closed film with control clarity located at a top limit of the Earth's troposphere (4 - 6 km). The film is supported at altitude by small additional atmospheric pressure and connected to ground by thin cables. It is known, the troposphere defines the Earth's weather. Authors show this closed dome allows to do a full control of the weather in a given region (the day is always fine, the rain is only in night, no strong wind). The average Earth (white cloudy) reflectance equal 0.3 - 0.5. That means the Earth losses about 30% - 50% of a solar energy. The dome controls the clarity of film and converts the cold regions to subtropics and creates the hot deserts, desolate wildernesses to the prosperous regions with temperate climate. That is a realistic and the cheapest method of the weather control in the Earth at the current time.

## Introduction

Governments spend billions of dollars to studying of weather. The many big government research scientific organizations and hundred thousands of scientists studying a Earth weather more then hundred years. There are gigantic numbers of scientific works about weather control. Most of them are out of practice. We cannot exactly predict weather at long period, to avert a rain, strong wind, storm, hurricane, tornado. We cannot control the clouds, temperature and humidity of atmosphere, power of rain. We cannot make better a winter and summer. We cannot convert a cold region to subtropics, a desolate wilderness to a prosperous region. We can only observe the storms and hurricanes and approximately predict their direction of movement. Every year the terrible storms, hurricanes, strong winds and rains, inundations destroy thousands of houses, kill thousands of men.

In this chapter, we consider a damage and prejudice from unnormal weather.

1. A **tropical cyclone (hurricane)** is a storm system fueled by the heat released when moist air rises and the water vapor in it condenses. The term describes the storm's origin in the tropics and its cyclonic nature, which means that its circulation is counterclockwise in the northern hemisphere and clockwise in the southern hemisphere. Tropical cyclones are distinguished from other cyclonic windstorms such as nor'easters, European windstorms, and polar lows by the heat mechanism that fuels them, which makes them "warm core" storm systems.

Depending on their location and strength, there are various terms by which tropical cyclones are known, such as hurricane, typhoon, tropical storm, cyclonic storm and tropical depression.

Tropical cyclones can produce extremely strong winds, tornadoes, torrential rain, high waves, and storm surges. The heavy rains and storm surges can produce extensive flooding. Although their effects on human populations can be devastating, tropical cyclones also can have beneficial effects by relieving drought conditions. They carry heat away from the tropics, an important mechanism of the global atmospheric circulation that maintains equilibrium in the earth's troposphere.

An average of 86 tropical cyclones of tropical storm intensity form annually worldwide, with 47 reaching hurricane/typhoon strength, and 20 becoming intense tropical cyclones (at least of Category 3 intensity).

Worldwide, tropical cyclone activity peaks in late summer when water temperatures are warmest. However, each particular basin has its own seasonal patterns. On a worldwide scale, May is the least active month, while September is the most active.

In the North Atlantic, a distinct hurricane season occurs from June 1 to November 30, sharply peaking from late August through September. The statistical peak of the North Atlantic hurricane season is September 10. The Northeast Pacific has a broader period of activity, but in a similar time frame to the Atlantic. The Northwest Pacific sees tropical cyclones year-round, with a minimum in February and a peak in early September. In the North Indian basin, storms are most common from April to December, with peaks in May and November.[22]

Table No. 1.

| Season Lengths and Seasonal Averages | | | | | |
|---|---|---|---|---|---|
| Basin | Season Start | Season End | Tropical Storms (>34 knots) | Tropical Cyclones (>63 knots) | Category 3+ Tropical Cyclones (>95 knots) |
| Northwest Pacific | – | – | 26.7 | 16.9 | 8.5 |
| South Indian | October | May | 20.6 | 10.3 | 4.3 |
| Northeast Pacific | May | November | 16.3 | 9.0 | 4.1 |
| North Atlantic | June | November | 10.6 | 5.9 | 2.0 |
| Australia Southwest Pacific | October | May | 10.6 | 4.8 | 1.9 |
| North Indian | April | December | 5.4 | 2.2 | 0.4 |

A mature tropical cyclone can release heat at a rate upwards of $6 \times 10^{14}$ watts.[3] Tropical cyclones on the open sea cause large waves, heavy rain, and high winds, disrupting international shipping and sometimes sinking ships. However, the most devastating effects of a tropical cyclone occur when they cross coastlines, making landfall. A tropical cyclone moving over land can do direct damage in four ways:

- **High winds** - Hurricane strength winds can damage or destroy vehicles, buildings, bridges, etc. High winds also turn loose debris into flying projectiles, making the outdoor environment even more dangerous.
- **Storm surge** - Tropical cyclones cause an increase in sea level, which can flood coastal communities. This is the worst effect, as historically cyclones claimed 80% of their victims when they first strike shore.
- **Heavy rain** - The thunderstorm activity in a tropical cyclone causes intense rainfall. Rivers and streams flood, roads become impassable, and landslides can occur. Inland areas are particularly vulnerable to freshwater flooding, due to residents not preparing adequately.

- **Tornado activity** - The broad rotation of a hurricane often spawns tornadoes. Also, tornadoes can be spawned as a result of eyewall mesovortices, which persist until landfall. While these tornadoes are normally not as strong as their non-tropical counterparts, they can still cause tremendous damage.[31]

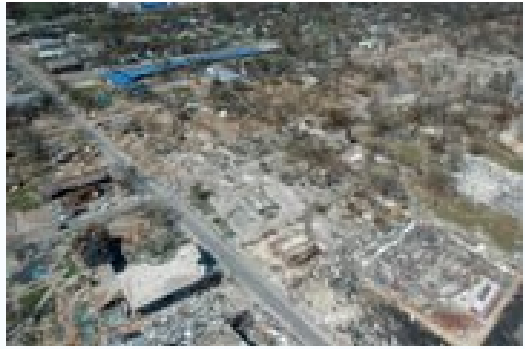

**Fig.1.** The aftermath of Hurricane Katrina in Gulfport, Mississippi. Katrina was the costliest tropical cyclone in United States history.

Often, the secondary effects of a tropical cyclone are equally damaging. These include:

- **Disease** - The wet environment in the aftermath of a tropical cyclone, combined with the destruction of sanitation facilities and a warm tropical climate, can induce epidemics of disease which claim lives long after the storm passes. One of the most common post-hurricane injuries is stepping on a nail in storm debris, leading to a risk of tetanus or other infection. Infections of cuts and bruises can be greatly amplified by wading in sewage-polluted water. Large areas of standing water caused by flooding also contribute to mosquito-borne illnesses.
- **Power outages** - Tropical cyclones often knock out power to tens or hundreds of thousands of people (or occasionally millions if a large urban area is affected), prohibiting vital communication and hampering rescue efforts.
- **Transportation difficulties** - Tropical cyclones often destroy key bridges, overpasses, and roads, complicating efforts to transport food, clean water, and medicine to the areas that need it.

. Hurricane Katrina is the most obvious example, as it devastated the region that had been revitalized after Hurricane Camille. Of course, many former residents and businesses do relocate to inland areas away from the threat of future hurricanes as well.

While the number of storms in the Atlantic has increased since 1995, there seems to be no signs of a numerical global trend; the annual global number of tropical cyclones remains about $90 \pm 10$. However, there is some evidence that the intensity of hurricanes is increasing. "Records of hurricane activity worldwide show an upswing of both the maximum wind speed in and the duration of hurricanes. The energy released by the average hurricane (again considering all hurricanes worldwide) seems to have increased by around 70% in the past 30 years or so, corresponding to about a 15% increase in the maximum wind speed and a 60% increase in storm lifetime."

Atlantic storms are certainly becoming more destructive financially, since five of the ten most expensive storms in United States history have occurred since 1990. This can be attributed to the increased intensity and duration of hurricanes striking North America and to the number of people living in susceptible coastal area following increased development in the region since the last surge in Atlantic hurricane activity in the 1960s.

Tropical cyclones that cause massive destruction are fortunately rare, but when they happen, they can cause damage in the range of billions of dollars and disrupt or end thousands of lives.

The deadliest tropical cyclone on record hit the densely populated Ganges Delta region of Bangladesh on November 13, 1970, likely as a Category 3 tropical cyclone. It killed an estimated 500,000 people. The North Indian basin has historically been the deadliest, with several storms since 1900 killing over 100,000 people, each in Bangladesh.

In the Atlantic basin, at least three storms have killed more than 10,000 people. Hurricane Mitch during the 1998 Atlantic hurricane season caused severe flooding and mudslides in Honduras, killing about 18,000 people and changing the landscape enough that entirely new maps of the country were needed. The Galveston Hurricane of 1900, which made landfall at Galveston, Texas as an estimated Category 4 storm, killed 8,000 to 12,000 people, and remains the deadliest natural disaster in the history of the United States. The deadliest Atlantic storm on record was the Great Hurricane of 1780, which killed about 22,000 people in the Antilles.

Hurricane Iniki in 1992 was the most powerful storm to strike Hawaii in recorded history, hitting Kauai as a Category 4 hurricane, killing six and causing $3 billion in damage. Other destructive Pacific hurricanes include Pauline and Kenna.

On March 26, 2004, Cyclone Catarina became the first recorded South Atlantic cyclone (cyclone is the southern hemispheric term for *hurricane*). Previous South Atlantic cyclones in 1991 and 2004 reached only tropical storm strength. Tropical cyclones may have formed there before 1960 but were not observed until weather satellites began monitoring the Earth's oceans in that year.

A tropical cyclone need not be particularly strong to cause memorable damage; Tropical Storm Thelma, in November 1991 killed thousands in the Philippines even though it never became a typhoon; the damage from Thelma was mostly due to flooding, not winds or storm surge. In 1982, the unnamed tropical depression that eventually became Hurricane Paul caused the deaths of around 1,000 people in Central America due to the effects of its rainfall. In addition, Hurricane Jeanne in 2004 caused the majority of its damage in Haiti, including approximately 3,000 deaths, while just a tropical depression.

On August 29, 2005, Hurricane Katrina made landfall in Louisiana and Mississippi. The U.S. National Hurricane Center, in its August review of the tropical storm season stated that Katrina was probably the worst natural disaster in U.S. history. Currently, its death toll is at least 1,836, mainly from flooding and the aftermath in New Orleans, Louisiana and the Mississippi Gulf Coast. It is also estimated to have caused $81.2 billion in property damage. Before Katrina, the costliest system in monetary terms had been 1992's Hurricane Andrew, which caused an estimated $39 billion (2005 USD) in damage in Florida.

**2**. A **flood (inundation)** is an overflow of water, an expanse of water submerging land, a deluge. In the sense of "flowing water", the word is applied to the inflow of the tide, as opposed to the outflow or "ebb". *The* Flood, the great Universal Deluge of myth and perhaps of history is treated at Deluge in mythology.

Since prehistoric times people have lived by the seas and rivers for the access to cheap and quick transportation and access to food sources and trade; without human populations near natural bodies of water, there would be no concern for floods. However fertile soil in a river delta is subject to regular inundation from normal variation in precipitation.

Floods from the sea can cause overflow or overtopping of flood-defenses like dikes as well as flattening of dunes or bluffs. Land behind the coastal defence may be inundated or experience damage. A flood from sea may be caused by a heavy storm (storm surge), a high tide, a tsunami, or a combination thereof. As many urban communities are located near the coast this is a major threat around the world.

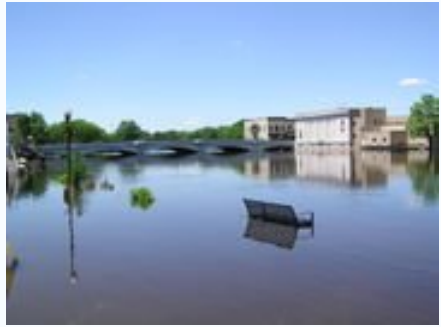

**Fig. 2.** Rock River floodwaters in downtown Fort Atkinson, Wisconsin.

Many rivers that flow over relatively flat land border on broad flood plains. When heavy the deposition of silt on the rich farmlands and can result in their eventual depletion. The annual cycle of flood and farming was of great significance to many early farming cultures, most famously to the ancient Egyptians of the Nile river and to the Mesopotamians of the Tigris and Euphrates rivers .

A flood happens when an area of land, usually low-lying, is covered with water. The worst floods usually occur when a river overflows its banks. An example of this is the January 1999 Queensland floods, which swamped south-eastern Queensland. Floods happen when soil and vegetation cannot absorb all the water. The water then runs off the land in quantities that cannot be carried in stream channels or kept in natural ponds or man-made reservoirs.

Periodic floods occur naturally on many rivers, forming an area known as the flood plain. These river floods usually result from heavy rain, sometimes combined with melting snow, which causes the rivers to overflow their banks. A flood that rises and falls rapidly with little or no advance warning is called a flash flood. Flash floods usually result from intense rainfall over a relatively small area. Coastal areas are occasionally flooded by high tides caused by severe winds on ocean surfaces, or by tidal waves caused by undersea earthquakes. There are often many causes for a flood.

Monsoon rainfalls can cause disastrous flooding in some equatorial countries, such as Bangladesh, Hurricanes have a number of different features which, together, can cause devastating flooding. One is the storm surge (sea flooding as much as 8 metres high) caused by the leading edge of the hurricane when it moves from sea to land. Another is the large amounts of precipitation associated with hurricanes. The eye of a hurricane has extremely low pressure, so sea level may rise a few metres in the eye of the storm. This type of coastal flooding occurs regularly in Bangladesh.

In Europe floods from sea may occur as a result from heavy Atlantic storms, pushing the water to the coast. Especially in combination with high tide this can be damaging.

Under some rare conditions associated with heat waves, flash floods from quickly melting mountain snow have caused loss of property and life.

Undersea earthquakes, eruptions of island volcanos that form a caldera, (such as Thera or Krakatau) and marine landslips on continental shelves may all engender a tidal wave called a tsunami that causes destruction to coastal areas. See the *tsunami* article for full details of these marine floods.

Floods are the most frequent type of disaster worldwide. Thus, it is often difficult or impossible to obtain insurance policies which cover destruction of property due to flooding, since floods are a relatively predictable risk.

- In 1983 the Pacific Northwest saw one of their worst winter floods. And the some of the Northwest states saw their wettest winter yet. The damage was estimated at 1.1 billion dollars.* In 1965 Hurricane Betsy flooded large areas of New Orleans for up to 10 days, drowning around 40 people.
- In 1957, storm surge flooding from Hurricane Audrey killed about 400 people in southwest Louisiana.
- The Hunter Valley floods of 1955 in New South Wales destroyed over 100 homes and caused 45,000 to be evacuated.
- The North Sea Flood of 1953 caused over 2,000 deaths in the Dutch province of Zeeland and the United Kingdom and led to the construction of the Delta Works and the Thames Barrier.
- The Lynmouth flood of 1952 killed only 34 people, however it was very destructive and destroyed over 80 buildings.
- The 1931 Huang He flood caused between 800,000 and 4,000,000 deaths in China, one of a series of disastrous floods on the Huang He.
- The Great Mississippi Flood in 1927 was one of the most destructive floods in United States history.

  The 2005 tragedy of New Orleans shows that disregard of protection of the USA's coastal cities (New York, Los Angles-San Pedro) from strong storm-caused waves, hurricane storm surges, and small tsunamis gives rise to gigantic damages, material losses, human deaths and injuries.

  The Metropolitan East Coast (MEC) region -- with New York City at its center -- has nearly 20 million people, a $1 trillion economy, and $2 trillion worth of built assets, nearly half of which are complex infrastructure.

  Many elements of transportation and other essential infrastructure systems in the MEC region, and even some of its regular building stock, are located at elevations from 6 to 20 feet above current sea level. This is well within the range of expected coastal storm surge elevation of 8 to more than 20 feet for tropical (hurricanes) and extra-tropical storms. Depending on which climate change scenarios apply, the sea level regional rise over the next 100 years will accelerate and amount to at most 3 feet by the year 2100. This seemingly modest increase in sea level has the effect to raise the frequency of coastal storm surges and related flooding by factors of 2 to 10, with an average of about 3.

  The rate of financial losses incurred from these coastal floods will increase accordingly. Expected annualized losses from coastal storms, already on the order of about $1 billion per year, would be small enough to be absorbed by the $1 trillion economy of the region. However, actual losses do not occur in regular annualized doses. Rather, they occur during infrequent extreme events that can cause losses of hundreds of billions of dollars for the largest events, albeit with low probability.

| Ten deadliest natural disasters | | | Table 2 |
|---|---|---|---|
| **Rank Event** | **Location** | **Date** | **Death Toll (Estimate)** |
| 1.  1931 Yellow River flood | Yellow River, China | Summer 1931 | 850,000-4,000,000 |
| 2.  1887 Yellow River flood | Yellow River, China | September-October 1887 | 900,000-2,000,000 |

| | | | | |
|---|---|---|---|---|
| 3. | 1970 Bhola cyclone | Ganges Delta, East Pakistan | November 13, 1970 | 500,000-1,000,000 |
| 4. | Earthquake | Eastern Mediterranean | 1201 | 1,000,000 |
| 5. | 1938 Yellow River flood | Yellow River, China | June 9th, 1938 | 500,000 - 900,000 |
| 6. | Shaanxi Earthquake | Shaanxi Province, China | January 23, 1556 | 830,000 |
| 7. | 2004 Indian Ocean earthquake/tsunami | Indian Ocean | December 26, 2004 | 225,000-275,000 |
| 8. | Tropical Cyclone | Haiphong, Vietnam | 1881 | 300,000 |
| 9. | Flood | Kaifeng, Henan Province, China | 1642 | 300,000 |
| 10. | Earthquake | Tangshan, China | July 28, 1976 | 242,000* |

* Official Government figure. Estimated death toll as high as 655,000.

**3. Brif information about cover film and liquid crystal.** Our dome cover (film) has 5 layers (fig. 4): transparant dielectric layer, conducting layer (about 1 - 3 μ), liquid crystal layer (about 10 - 100 μ), conducting layer (for example, $SnO_2$), and transparant dielectric layer. Common thickness is 0.1 - 0.5 mm. Control voltage is 5 - 10 V. Film is produced the industry and it not expensive.

Liquid crystals (LC) are substances that exhibit a phase of matter that has properties between those of a conventional liquid, and those of a solid crystal.

Liquid crystals find wide use in liquid crystal displays (LCD), which rely on the optical properties of certain liquid crystalline molecules in the presence or absence of an electric field. The electric field can be used to make a pixel switch between clear or dark on command. Color LCD systems use the same technique, with color filters used to generate red, green, and blue pixels. Similar principles can be used to make other liquid crystal based optical devices. Liquid crystal in fluid form is used to detect electrically generated hot spots for failure analysis in the semiconductor industry. Liquid crystal memory units with extensive capacity were used in Space Shuttle navigation equipment. It is also worth noting that many common fluids are in fact liquid crystals. Soap, for instance, is a liquid crystal, and forms a variety of LC phases depending on its concentration in water.

The conventional control clarity film reflected a superfluos energy back to space. If film has solar cells that converts the superfluos solar energy into electricity.

## 2. DESCRIPTION AND INNOVATIONS

Our idea is a dome covering a big region (city, large bad area, country, continent) by a thin film with control clarity (reflectivity, carrying capacity of solar spectrum). The film is located at high altitude (4 - 6 km) which include the Earth's troposphere where are the main climatic changes. The film is support at the altitude by a small additional air pressure produced by ground ventilators and connected to Earth's ground by cables. The closed area is also divided by sub-areas by film having control clarity. That allows to make different conditions (solar heating) in sub-areas and pumping hot, warm, cold, moist air from one sub-area to other sub-area. There are a cheap film having liquid crystal and conducting layers. The clarity of them is controlled by electric voltage. They can pass or blockade the solar light (or parts of solar spectrum) and pass or blockade the Earth radiation. The outer and incite radiations have different wave lengths. That makes to control of them separately and to control a heating of the Earth surface. In conventional conditions about 50% of the solar energy reaches the Earth surface. The most part is reflected back to outer space by the white clouds. In our closed system the clouts (and rain) will be made in a night when temperature is low. That means the many cold regions (Alaska, Siberia) may be accepted more solar energy and became a temperate climate or sub-tropic climate. That also means the Sahara desert can be a prosperous area with fine climate and with closed-loop water cycle.

The building of film dome is very ease. We spread out the film over Earth surface, turn on the pumping propellers and film is risen by air to needed altitude limited by the support cables. The bid damage of film is not trouble because the additional air pressure is very small and air leakage is compensated by propeller pumps.

The other advantages of the suggested method is possibility to pain the pictures on sky (dome), to show films on the sky by projector, to suspend illuminations, decorations, and air tramway. The long distance aircraft fly at altitude 8 - 11 km and our dome do not trouble for it. The support cable will have illumination and internal helicopters also will avoid the contact with them.

The people throw out hundreds the thin film plastic bags from purchases every month. If we will collect them and use for the offered dome, we make fine our weather, get new territory for living with wonderful climate.

Our design for the dome is presented in Fig. 3, which includes the thin inflated film dome.  The innovations are listed here: (1) the construction is air-inflatable; (2) each dome is fabricated with very thin, transparent film (thickness is 0.1 to 0.3 mm) having the control clarity without rigid supports; (3) the enclosing film has two conductivity layers plus a liquid crystal layer between them which changes its clarity, color and reflectivity under an electric voltage (fig,4); (4) the bound section of dome has a hemisphere form (#5, fig.3) . The air pressure is more in these sections and they protect the central sections from outer wind.

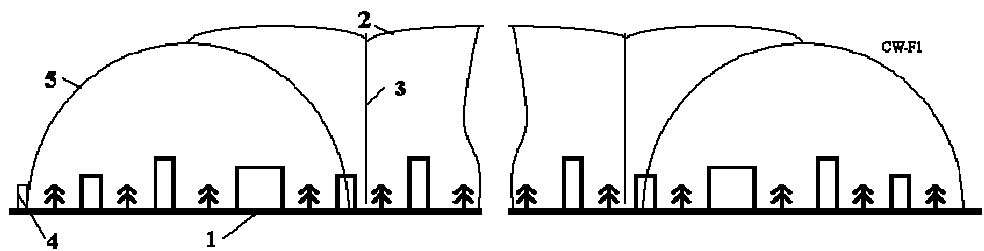

**Fig.3**. Film dome over city. Notations: 1 - city, 2 - thin film cover with control clarity, 3 - support cable, 4 - exits and ventilators, 5 - semi-cylindrical border section.

Fig. 3 illustrates the thin transparent control dome cover we envision.  The inflated textile shell—technical "textiles" can be woven or non-woven (films)—embodies the innovations listed: (1) the film is very thin, approximately 0.1 to 0.3 mm.  A film this thin has never before been used in a major building; (2) the film has two strong nets, with a mesh of about $0.1 \times 0.1$ m and $a = 1 \times 1$ m, the threads are about 0.5 mm for a small mesh and about 1 mm for a big mesh.  The net prevents the watertight and airtight film covering from being damaged by vibration; (3) the film incorporates a tiny electrically conductive wire net with a mesh about 0.1 x 0.1 m and a line width of about 100 μ and a thickness near 10 μ.  The wire net is electric (voltage) control conductor. It can inform the dome supervisors concerning the place and size of film damage (tears, rips, etc.) ; (4) the film may be twin-layered with the gap — $c = 1$ m and $b = 2$ m—between covering's layers for heat saving. In polar regions this multi-layered covering is the main means for heat insulation and puncture of one of the layers wont cause a loss of shape because the film's second layer is unaffected by holing; (5)

the airspace in the dome's covering can be partitioned, either hermetically or not; and (6) part of the covering can have a very thin shiny aluminum coating that is about 1μ for reflection of unnecessary solar radiation in equatorial or polar regions [1].

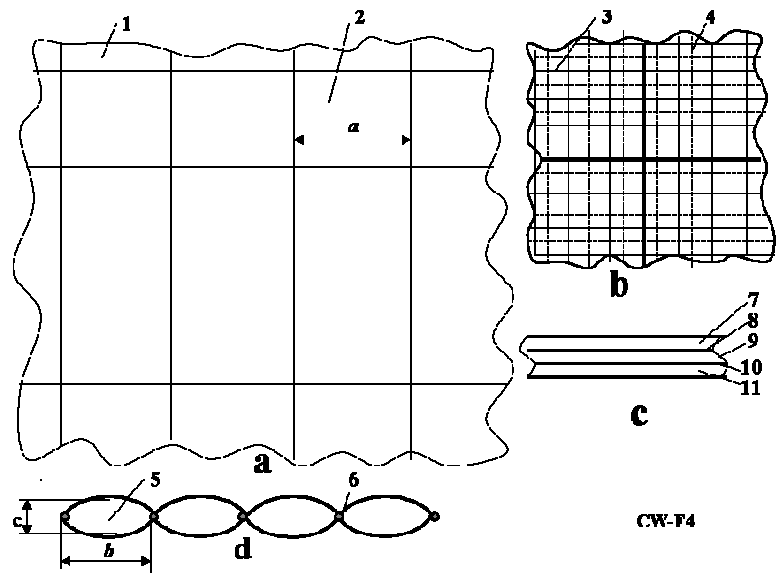

**Fig.4.** Design of covering membrane. Notations: (a) Big fragment of cover with control clarity (reflectivity, carrying capacity); (b) Small fragment of cover; (c) Cross-section of cover (film)having 5 layers; (d) Longitudinal cross-section of cover for cold regions; 1 - cover; 2 -mesh; 3 - small mesh; 4 - thin electric net; 5 - cell of cover; 6 - tubes;: 7 - transparent dielectric layer, 8 - conducting layer (about 1 - 3 μ), 9 - liquid crystal layer (about 10 - 100 μ), 10 - conducting layer, and 11 - transparent dielectric layer. Common thickness is 0.1 - 0.5 mm. Control voltage is 5 - 10 V.

## 3. THEORY AND COMPUTATIONS DOME

As wind flows over and around a fully exposed, nearly completely sealed inflated dome, the weather affecting the external film on the windward side must endure positive air pressures as the wind stagnates. Simultaneously, low air pressure eddies will be present on the leeward side of the dome. In other words, air pressure gradients caused by air density differences on different parts of the dome's envelope is characterized as the "buoyancy effect". The buoyancy effect will be greatest during the coldest weather when the dome is heated and the temperature difference between its interior and exterior are greatest. In extremely cold climates such as the Arctic and Antarctic Regions the buoyancy effect tends to dominate dome pressurization.

Our basic computed equations, below, are derived from a Russian-language textbook. Solar radiation impinging the orbiting Earth is approximately 1400 W/m$^2$. The average Earth reflection by clouds and the sub-aerial surfaces (water, ice and land) is about 0.3. The Earth-atmosphere adsorbs about 0.2 of the Sun's radiation. That means about $q_0$ = 700 W/m$^2$s of solar energy (heat) reaches our planet's surface at the Equator. Our troposphere dome does not have clouds in light time and contains about 1/3 part of Earth atmosphere. That means we can adsorb about 70 - 80% of solar energy. It is useful for polar regions and in winter time.

The solar spectrum is graphed in Fig. 5.

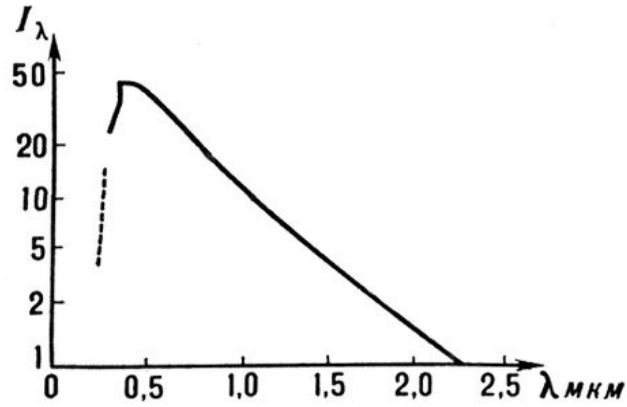

**Fig.5.** Spectrum of solar radiation. Visible light is 0,4 - 0,8 $\mu$.

The visible part of the Sun's spectrum is only $\lambda$ = 0.4 to 0.8 $\mu$.. Any warm body emits radiation. The emission wavelength depends on the body's temperature. The wavelength of the maximum intensity (see Fig. 5) is governed by the black-body law originated by Max Planck (1858-1947):

$$\lambda_m = \frac{2.9}{T}, \quad [mm], \tag{1}$$

where $T$ is body temperature, $^{o}$K. For example, if a body has an ideal temperature 20 $^{o}$C ($T$ = 293 $^{o}$K), the wavelength is $\lambda_m$ = 9.9 $\mu$.

The energy emitted by a body may be computed by employment of the Josef Stefan-Ludwig Boltzmann law.

$$E = \varepsilon \sigma_s T^4, \quad [W/m^2], \tag{2}$$

where $\varepsilon$ is coefficient of body blackness ($\varepsilon$ =0.03 $\div$ 0.99 for real bodies), $\sigma_s$ = 5.67×10 $^{-8}$ [W/m$^2$ ·K] Stefan-Boltzmann constant. For example, the absolute black-body ($\varepsilon$ = 1) emits (at $T$ = 293 $^{0}$K) the energy $E$ = 418 W/m$^2$.

Amount of the maximum solar heat flow at 1 m$^2$ per 1 second of Earth surface is

$$q = q_o \cos{(\varphi \pm \theta)} \quad [W/m^2], \tag{3}$$

where $\varphi$ is Earth longevity, $\theta$ is angle between projection of Earth polar axis to the plate which is perpendicular to the ecliptic plate and contains the line Sun-Earth and the perpendicular to ecliptic plate. The sign "+" signifies Summer and the "-" signifies Winter, $q_o \approx$ 700 W/m$^2$ is the annual average solar heat flow to Earth at equator corrected for Earth reflectance. For our case this magnitude can reach $q_o \approx$ 1000 - 1100 W/m$^2$.

This angle is changed during a year and may be estimated for Earth's North Polar Region hemisphere by the following the first approximation equation:

$$\theta = \theta_m \cos{\omega}, \quad \text{where} \quad \omega = 2\pi \frac{N}{364}, \tag{4}$$

where $\theta_m$ is maximum $\theta$, $|\theta_m|$ = 23.5$^{o}$ =0.41 radian; $N$ is number of day in a year. The computations for Summer and Winter are presented in fig.6.

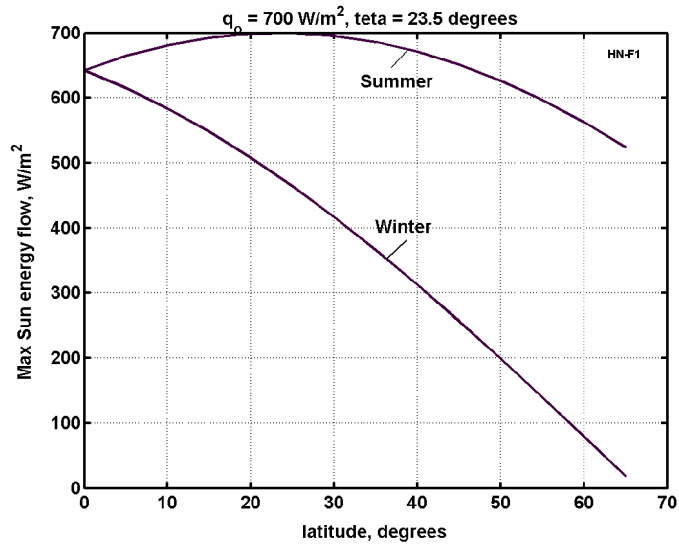

**Fig.6.** Maximum Sun radiation flow at Earth surface via Earth latitude and season without dome.

The heat flow for a hemisphere having reflector [1] at noon may be computed by equation

$$q = c_1 q_0 \left[ \cos(\varphi - \theta) + S \sin(\varphi + \theta) \right], \tag{5}$$

where $S$ is fraction (relative) area of reflector to service area of "Evergreen" dome [1]. For reflector of Fig.1 [1] $S = 0.5$; $c_1$ is film transparency coefficient ($c_1 \approx 0.9 - 0.95$).

The daily average solar irradiation (energy) is calculated by equation

$$Q = 86400\, c\, q t, \quad \text{where} \quad t = 0.5 \left( 1 + \tan \varphi \tan \theta \right), \quad \left| \tan \varphi \tan \theta \right| \leq 1, \tag{6}$$

where $c$ is daily average heat flow coefficient, $c \approx 0.5$ without dome, $c \approx 0.75$ with dome; $t$ is relative daily light time, $86400 = 24 \times 60 \times 60$ is number of seconds in a day.

The computation for relative daily light period is presented in Fig. 7.

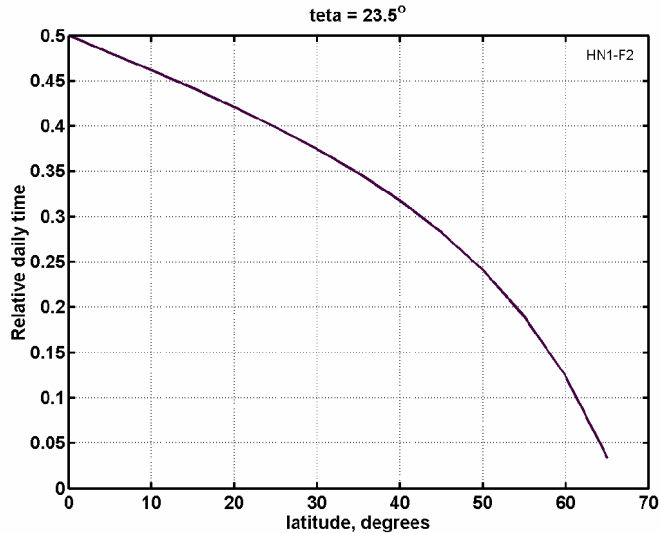

**Fig.7**. Relative daily light time via Earth latitude.

The heat loss flow per 1 m$^2$ of dome film cover by convection and heat conduction is (see [2]):

$$q = k(t_1 - t_2), \quad \text{where} \quad k = \frac{1}{1/\alpha_1 + \sum_i \delta_i / \lambda_i + 1/\alpha_2}, \tag{7}$$

where $k$ is heat transfer coefficient, W/m$^2$K; $t_{1,2}$ are temperatures of the inter and outer multi-layers of the heat insulators, $^{\circ}$C; $\alpha_{1,2}$ are convention coefficients of the inter and outer multi-layers of heat insulators ($\alpha = 30 \div 100$), W/m$^2$K; $\delta_i$ are thickness of insulator layers; $\lambda_i$ are coefficients of heat transfer of insulator layers (see Table 1), m; $t_{1,2}$ are temperatures of initial and final layers $^{\circ}$C.

The radiation heat flow per 1 m$^2$s of the service area computed by equations (2):

$$q = C_r \left[ \left( \frac{T_1}{100} \right)^4 - \left( \frac{T_2}{100} \right)^4 \right], \quad \text{where} \quad C_r = \frac{c_s}{1/\varepsilon_1 + 1/\varepsilon_2 - 1}, \quad c_s = 5.67 \ [\text{W/m}^2\text{K}^4], \quad (8)$$

where $C_r$ is general radiation coefficient, $\varepsilon$ are black body rate (emittance) of plates (see Table 2); $T$ is temperatures of plates, $^{\text{o}}$K.

The radiation flow across a set of the heat reflector plates is computed by equation

$$q = 0.5 \frac{C_r'}{C_r} q_r, \qquad (9)$$

where $C_r'$ is computed by equation (8) between plate and reflector.

The data of some construction materials is found in Table 3, 4.

**Table 3**. [2], p.331. Heat Transferring.

| Material | Density, kg/m$^3$ | Thermal conductivity, $\lambda$, W/m· $^{\text{o}}$C | Heat capacity, kJ/kg. $^{\text{o}}$C |
|----------|----------|----------|----------|
| Concrete | 2300 | 1.279 | 1.13 |
| Baked brick | 1800 | 0.758 | 0.879 |
| Ice | 920 | 2.25 | 2.26 |
| Snow | 560 | 0.465 | 2.09 |
| Glass | 2500 | 0.744 | 0.67 |
| Steel | 7900 | 45 | 0.461 |
| Air | 1.225 | 0.0244 | 1 |

As the reader will see, the air layer is the best heat insulator. We do not limit its thickness $\delta$.

**Table 4**. [2], p. 465. Emittance

| Material | Emittance, $\varepsilon$ | Material | Emittance, $\varepsilon$ | Material | Emittance, $\varepsilon$ |
|----------|----------|----------|----------|----------|----------|
| Bright Aluminum $t = 50 \div 500 \ ^{\text{o}}$C | 0.04 - 0.06 | Baked brick $t = 20 \ ^{\text{o}}$C | 0.88 - 0.93 | Glass $t = 20 \div 100 \ ^{\text{o}}$C | 0.91 - 0.94 |

As the reader will notice, the shiny aluminum louver coating is excellent mean jalousie against radiation losses from the dome.

The general radiation heat $Q$ computes by equation [6]. Equations [1] – [9] allow computation of the heat balance and comparison of incoming heat (gain) and outgoing heat (loss).

The computations of heat balance of a dome of any size in the coldest wintertime of the Polar Regions are presented in Fig. 8.

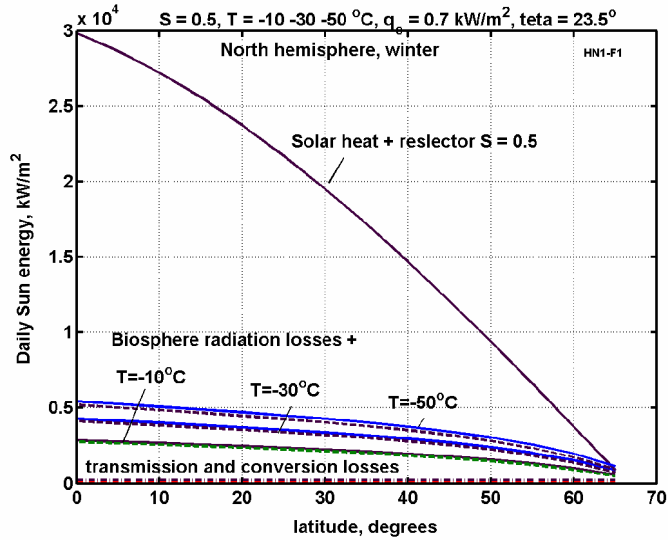

**Fig. 8.** Daily heat balance through 1 m² of dome during coldest winter day versus Earth's latitude (North hemisphere example). Data used for computations (see Eq. (1) - (9)): temperature inside of dome is $t_1$= +20 °C, outside are $t_2$ = -10, -30, -50 °C; reflectivity coefficient of mirror is $c_2$= 0.9; coefficient transparency of film is $c_1$ = 0.9; convectively coefficients are $\alpha_1 = \alpha_2$ = 30; thickness of film layers are $\delta_1 = \delta_2$ =0.0001 m; thickness of air layer is $\delta$ = 1 m; coefficient of film heat transfer is $\lambda_1 = \lambda_3$ = 0.75, for air $\lambda_2$ = 0.0244; ratio of cover blackness $\varepsilon_1 = \varepsilon_3$ = 0.9, for louvers $\varepsilon_2$ = 0.05.

The thickness of the dome envelope, its sheltering shell of film, is computed by formulas (from equation for tensile strength):

$$\delta_1 = \frac{Rp}{2\sigma}, \quad \delta_2 = \frac{Rp}{\sigma}, \tag{10}$$

where $\delta_1$ is the film thickness for a spherical dome, m; $\delta_2$ is the film thickness for a cylindrical dome, m; $R$ is radius of dome, m; $p$ is additional pressure into the dome, N/m²; $\sigma$ is safety tensile stress of film, N/m².

The dynamic pressure from wind is

$$p_w = \frac{\rho V^2}{2}, \tag{11}$$

where $\rho$ = 1.225 kg/m³ is air density; $V$ is wind speed, m/s.

For example, a storm wind with speed $V$ = 20 m/s, standard air density is $\rho$ = 1.225 kg/m³. Then dynamic pressure is $p_w$ = 245 N/m². That is four time less when internal pressure $p$ = 1000 N/m². When the need arises, sometimes the internal pressure can be voluntarily decreased, bled off.

In Fig. 8 the alert reader has noticed: the daily heat loss is about the solar heat in the very coldest Winter day when a dome located above 60⁰ North or South Latitude and the outside air temperature is −50 ⁰C.

In [1] we show the heat loss of the dome in Polar region is less than 14 times the heat of the buildings inside unprotected by an inflated dome.

We consider a two-layer dome film and one heat screen. If needed, better protection can further reduce the head losses as we can utilize inflated dome covers with more layers and more heat screens. One heat screen decreases heat losses by 2, two screens can decrease heat flow by 3 times, three by 4 times, and so on. If the Polar Region domes have a mesh structure, the heat transfer decreases proportional to the summary thickness of its enveloping film layers.

The dome shelter innovations outlined here can be practically applied to many climatic regimes (from Polar to Tropical). The North and South Poles may, during the 21ˢᵗ Century, become places of cargo and passenger congregation since the a Cable Space Transportation System, installed on Antarctica's ice-cap and on a floating artificial ice island has been proposed the would transfer people and things to and from the Moon.[i]

## 4. DISCUSSION

As with any innovative macro-project proposal, the reader will naturally have many questions. We offer brief answers to the four most obvious questions our readers are likely to ponder.

(1)   *How can snow and ice be removed from the dome?*

The rain, snow clouds located in Earth troposphere lower 4 km altitude. Our dome has height 4 - 6 km. If water appears over film, it flows down through special opening. If snow appears over film, the control made the black film, the sun flux the snow.  The film cover is flexible and has a lift force of about 1 -100 kg/m$^2$. We imagine that a controlled change of interior air pressure will serve to shake the snow and ice off.

(2) *Storm wind.*

The storm wind can be only on bounding sections of dome. They are special semi-cylindrical form (fig.3) and more strong then central sections.

(3) *Cover damage.*

The envelope contains a cable mesh so that the film cannot be damaged greatly.  Electronic signals alert supervising personnel of any rupture problems.

(3) *What is the design life of the film covering?*

Depending on the kind of materials used, it may be as much a decade.  In all or in part, the cover can be replaced periodically.

## 5. CONCLUSION

The control of Regional and Global Earth Weather is important problem of humanity. That dramatically increases the territory suitable for men living, sown area, crop capacity. That allows to convert all Earth lend such as Alaska, North Canada, Siberia, deserts Sahara or Gobi in prosperous garden. The suggested method is very cheap (cost of covering 1 m$^2$ is about 2 - 15 cents) and may be utilized at present time. We can start from small area, from small towns in bad regions and extended in large area.

Film domes can foster the fuller economic development of cold regions such as the Earth's Arctic and Antarctic and, thus, increase the effective area of territory dominated by humans. Normal human health can be maintained by ingestion of locally grown fresh vegetables and healthful "outdoor" exercise.  The domes can also be used in the Tropics and Temperate Zone. Eventually, they may find application on the Moon or Mars since a vertical variant, inflatable towers to outer space, are soon to become available for launching spacecraft inexpensively into Earth-orbit or interplanetary flights.

The closed problems are researched in references [3]-[10].

## REFERENCES

1. Bolonkin, A.A. and R.B. Cathcart, Inflatable 'Evergreen' Dome Settlements for Earth's Polar Regions. Clean. Techn. Environ. Policy. DOI 10.1007/s10098.006-0073.4 .
2. Naschekin, V.V., *Technical thermodynamic and heat transmission*. Public House High University, Moscow. 1969 (in Russian).
3. Bolonkin, A.A. and R.B. Cathcart, "A Cable Space Transportation System at the Earth's Poles to Support Exploitation of the Moon", *Journal of the British Interplanetary Society* 59: 375-380, 2006.
4. Bolonkin A.A., Cheap Textile Dam Protection of Seaport Cities against Hurricane Storm Surge Waves, Tsunamis, and Other Weather-Related Floods, 2006. http://arxiv.org.
5. Bolonkin, A.A. and R.B. Cathcart, Antarctica: A Southern Hemisphere Windpower Station? Arxiv, 2007.
6. Cathcart R.B. and Bolonkin, A.A. Ocean Terracing, 2006. http://arxiv.org.
7. Bolonkin, A.A. and R.B. Cathcart, The Java-Sumatra Aerial Mega-Tramway, 2006. http://arxiv.org.
8. Bolonkin, A.A., "Optimal Inflatable Space Towers with 3-100 km Height", *Journal of the British Interplanetary Society* Vol. 56, pp. 87 - 97, 2003.
9. Bolonkin A.A., *Non-Rocket Space Launch and Flight, Elsevier*, London, 2006, 488 ps.
10. Macro-Engineering: *A Challenge for the Future*. Springer, 2006. 318 ps. Collection articles.
11. Bolonkin A.A., Cathcart R.B., Inflatable 'Evergreen' Polar Zone Dome (EPZD) Settlements, 2006. http://arxiv.org.